\documentclass[pra,preprint,floatfix,onecolumn,superscriptaddress]{revtex4-1}
\usepackage[dvips]{graphicx}
\usepackage[english]{babel}
\usepackage{amsmath}
\usepackage{amssymb}
\usepackage{tensor}
\usepackage{subfigure}

\newcommand{\vk}{\mathbf{k}}
\newcommand{\vn}{\mathbf{n}}
\newcommand{\vbr}{\mathbf{r}}
\newcommand{\be}{\begin{eqnarray}}
\newcommand{\ee}{\end{eqnarray}}
\newcommand{\p}{\partial}
\newcommand{\hx}{\hat{x}}

\newcommand{\dc}{c^{\dagger}}

\def\ket#1{|#1\rangle}
\def\bra#1{\langle #1 |}

\begin{document}

\title{Disorder effects on the Topological Superconductor with Hubbard Interactions}
\author{Yiting Deng}
\affiliation{College of Physics, Sichuan University, Chengdu, Sichuan 610064, China}
\email{heyan$_$ctp@scu.edu.cn}

\author{Yan He}
\affiliation{College of Physics, Sichuan University, Chengdu, Sichuan 610064, China}
\email{heyan$_$ctp@scu.edu.cn}

\begin{abstract}
We study the two-dimensional disordered topological superconductor with Hubbard interactions. When the magnitude of the pairing potential is tuned to special values, this interacting model is exactly solvable even when disorders are imposed on the potential term or coupling constants. The topology of this model is investigated in detail by the real space Chern number formula, which computes the topological index of disordered systems to high precisions. It is found that the disorders can drive the system from topological trivial phase to a non-trivial phase, which generalizes the topological Anderson phenomena to interacting models.
The self-consistent Born approximation is also employed to understand the influence of the disorders on the parameters of the interacting topological superconductor. It provide an alternative way to understand the topological transitions at weak disordered region.
\end{abstract}

\maketitle

\section{Introduction}

In the past decade, there are tremendous progress that have been achieved in the area of the topological properties of materials \cite{Kane-review,Zhang-review}. However, most of these studies are restricted to the free fermion systems or weakly interacting systems. It is well known that the understanding of strongly correlated systems such as Hubbard model \cite{arovas22,hirsch85,schulz90,Fradkin-book} is very difficult since the perturbation theory is not reliable for these systems due to the lack of small expansion parameters. Recently, there emerges a series of topological models with Hubbard interactions which are exact solvable due to the existence of flat bands. This type of models are partly inspired by the Kitaev model \cite{Kitaev-model} on the honeycomb lattice and are extensively studied by several authors \cite{TKNg2018:prl,TKNg2019:prb,miao17,miao19,Ezawa18}. They provide some interesting examples for understanding how the interactions affect the topological properties. Along this line, we also proposed an exactly solvable topological superconductor with Hubbard interaction \cite{deng21}, which enriches this class of models.

In the real materials, disorders or impurities are un-avoidable. In view of this, the studies of the disordered systems are also an important topic in condensed matter physics. A classical example of disorder effects is the Anderson localization \cite{Anderson79} which induces metal-insulator transitions in free fermion systems. An intriguing question is that how the disorders will affect the exact solvable interacting topological models we mentioned above. It is well-known that the topological phase can persist until the disorder strength reaches some critical values. After that, the system will return to the trivial phase. However, it is also found that the disorders can induce nontrivial topology in the system that is trivial in the clean limit. Ever since the pioneer work of the so-called topological Anderson insulator \cite{Li09,shen-book} has been carried out, many other theoretical models have been proposed to understand these disorder induced topological phases \cite{chenrui2017PRB,chen2017disorderPRB,klier2019weak,Guo-1,Guo-2,Hughes,Xie}. The disorder induced topological phase transitions were also discussed in works \cite{Louvet_2016PRB,Ryu_2012PRB}. In this paper, we will investigate how the disorders influence the topology of our interacting topological superconductor. The exact solvability allows us to perform a detailed study of the interplay between the interactions and disorders. Furthermore, it is also possible to make reliable calculation when the disorders are imposed on the coupling constants.

The rest of this paper is organized as follows. In section \ref{model}, we present the TS model with Hubbard interactions and show that this model is exactly solvable when flat band condition is satisfied. In section \ref{method}, we introduce the real space Chern number formula which is suitable for investigations of the topology of disordered systems. The self-consistent Born approximation is also discussed, which can provide more insights about the influence of disorders. The numerical results of the disordered interacting TS model based on the above mentioned two methods are presented in section \ref{num}. Finally, we briefly conclude in section \ref{conclusion}.

\section{The topological superconductor with Hubbard interaction}
\label{model}

In this section, we introduce the model of topological superconductor which can support flat bands with a suitable choice of parameters. Owing to the presence of flat bands, this type of model is exactly solvable when a Hubbard interaction term is included \cite{miao17,Ezawa18}. The Hamiltonian of the topological superconductor (TS) in momentum space can be expressed in the conventional Bogoliubov-de Gennes form as \cite{deng21}
\be
\mathcal{H}_{TS}=\sum_{\vk}\sum_{a,b}\psi_{a,\mathbb{\mathbf{k}}}^{\dagger}(H_{TS}(\vk)_{ab}\psi_{b,\mathbf{k}},\quad
H_{TS}(\vk)=\left(\begin{array}{cc}
\xi(\mathbf{k}) & \Delta\phi(\mathbf{k})\\
\Delta\phi^{\dagger}(\mathbf{k}) & -\xi(\mathbf{-k})^{*}
\end{array}\right)
\label{ham:eq1}
\ee
Here we have defined $\psi_{\mathbf{k}}=(c_{1,\vk}, c_{2,\vk}, c_{1,-\vk}^{\dagger}, c_{2,-\vk}^{\dagger})^{T}$. The first sub-indices denote the two different orbitals on each site. The lattice momentum $\mathbf{k}$ takes values from the Brillouin zone of a square lattice. The hopping and pairing terms are given by
\begin{equation}\begin{aligned}
\xi(\mathbf{k})=\sum_{j=1}^{3}R_{j}\sigma_{j},\quad
\phi(\mathbf{k})=i(R_{1}\sigma_{2}-R_{2}\sigma_{1})-R_{3}\sigma_{0}
\label{ene:psi}
\end{aligned}\end{equation}
Where $\sigma_{j}$ with $j=1,2,3$ are Pauli matrices and $\sigma_0$ is 2 by 2 identity matrix. The 3 coefficients $R_{j}$ are chosen as follows
\begin{equation}\begin{aligned}
R_{1}=m+\cos k_{x}+\cos k_{y},\quad R_{2}=\sin k_{x},\quad R_{3}=\sin k_{y}
\label{par:eq3}
\end{aligned}\end{equation}
One can see that the hopping term alone is equivalent to the famous Qi-Wu-Zhang model of Chern insulator \cite{QWZ}.

It is also convenient to rewrite the topological superconductor model as
\be
&&H_{\textrm{TS}}=H_{\textrm{CI}}+H_{\textrm{pair}},\\
&&H_{\textrm{CI}}=R_1\Gamma_{31}+R_2\Gamma_{32}+R_3\Gamma_{03},\\
&&H_{\textrm{pair}}=-\Delta(R_1\Gamma_{22}-R_2\Gamma_{21}+R_3\Gamma_{10})
\label{Ga}
\ee
Here we defined $\Gamma_{ij}=\tau_i\otimes\sigma_j$ with Pauli matrices $\tau_i$ and $\sigma_j$ acting on the Nambu spinor space and the orbital space respectively. One can see that the 4 energy bands of $H_{TS}$ is given by
\be
E=\pm(\Delta\pm1)\sqrt{R_1^2+R_2^2+R_3^2}
\ee
Therefore, when $\Delta=\pm1$, there will appear two flat bands $E=0$. Note that the pairing term is chosen to share the same structure as the hopping term. This is the reason why the $H_{TS}$ can support two flat bands with special choice of parameters.

The topological properties of the model with Eq.(\ref{ham:eq1}) is the same as the Qi-Wu-Zhang model. One find that the Chern number $C=\pm1$ when $|m|<2$ and $C=0$ otherwise. When $C=\pm1$, the above model is topologically nontrivial, which also indicates that chiral edge modes will appear if we impose an open boundary condition to the system. One the other hand, for $C=0$, the system is topologically trivial with no edge modes.

Up to now, the above topological superconductor is a model of spinless fermions. It is straightforward to include the spin degree of freedom which gives rise to the Hamiltonian of the first quantization form as
\begin{equation}\begin{aligned}
H=H_{TS}\otimes s_{0}
\label{ham:eq4}
\end{aligned}\end{equation}
Here $s_{0}$ is a $2\times2$ identity matrix acting on the spinor space and $H_{TS}$ is from Eq. (\ref{ham:eq1}). Now we can define the Hubbard interaction as the on-site repulsion between spin up and spin down fermions as follows
\begin{equation}\begin{aligned}
H_{int}=U\sum_{\mathbf{n},j}(c_{\mathbf{n},j,\uparrow}^{\dagger}c_{\mathbf{n},j,\uparrow}-\frac{1}{2})
(c_{\mathbf{n},j,\downarrow}^{\dagger}c_{\mathbf{n},j,\downarrow}-\frac{1}{2})
\label{ham:hub}
\end{aligned}\end{equation}
Since the Hubbard interaction term is expressed in the real space, it is more convenient to transfer the hopping and pairing terms of Eq.(\ref{ham:eq4}) to the real space. In the second quantization form, The resulting $H_{\textrm{CI}}$ and $H_{\textrm{pair}}$ terms can be written as
\be
&&H_{\textrm{CI}}=2\sum_{\vn,s}\Big(\dc_{\vn,s}\frac{\sigma_1-i\sigma_2}{2}c_{\vn+\hat{x},s}
+\dc_{\vn,s}\frac{\sigma_1-i\sigma_3}{2}c_{\vn+\hat{y},s}+h.c.\Big)
+2\sum_{\vn,s} m \dc_{\vn,s} \sigma_1 c_{\vn,s}\label{CI}\\
&&H_{\textrm{pair}}=2\Delta\sum_{\vn,s}\Big(\dc_{\vn,s}\frac{i\sigma_2-\sigma_1}2\dc_{\vn+\hat{x},s}
+\dc_{\vn,s}\frac{i\sigma_2+i\sigma_0}2\dc_{\vn+\hat{y},s}+h.c.\Big)\nonumber\\
&&\qquad+\Delta\sum_{\vn,s}\Big(m \dc_{\vn,s}(i\sigma_2) \dc_{\vn,s}+h.c.\Big)\label{pair}
\ee
Here $\vn=(n_x,n_y)$ labels the lattice site on a square lattice and $s=\uparrow,\downarrow$ labels the spin up and down. The two-component fermion operator is defined as $c_{\vn,s}=(c_{\vn,1,s}, c_{\vn,2,s})^T$, where the second index labels the two orbitals. We also introduce $\hat{x}$ and $\hat{y}$ representing the unit vector along the $x$ and $y$ direction respectively. Now the full interacting topological superconductor can be summarized as
\be
H_{\textrm{full}}=H_{\textrm{CI}}+H_{\textrm{pair}}+H_{\textrm{int}}
\label{H-full}
\ee

In order to understand the reason why the above interacting model is exactly solvable, we introduce the Majorana fermion operators as follows
\be
&&c_{\mathbf{n},1,s}=a_{\mathbf{n},1,s}+ib_{\mathbf{n},1,s},\quad
c_{\mathbf{n},1,s}^{\dagger}=a_{\mathbf{n},1,s}-ib_{\mathbf{n},1,s}\label{ma:c1}\\
&&c_{\mathbf{n},2,s}=b_{\mathbf{n},2,s}+ia_{\mathbf{n},2,s},\quad
c_{\mathbf{n},2,s}^{\dagger}=b_{\mathbf{n},2,s}-ia_{\mathbf{n},2,s}
\label{ma:c2}
\ee
If the Majorana fermion operators is substituted into the hopping and pairing terms of Eq.(\ref{CI}) and (\ref{pair}), we find that the terms involving $a_{\vn,j,s}$ are all vanished when the flat band condition $\Delta=1$ is satisfied (Detailed calculations can be found in \cite{deng21}). In terms of the Majorana fermions, the Hubbard interaction can be expressed as
\be
H_{\textrm{int}}=U\sum_{\vn,j}(2ia_{\vn,j,\uparrow}b_{\vn,j,\uparrow})(2ia_{\vn,j,\downarrow}b_{\vn,j,\downarrow})
\ee
If we define $D_{\vn,j}=4ia_{\vn,j,\uparrow}a_{\vn,j,\downarrow}$, it is easy to see that $[D_{\vn,j},H]=0$ in all sites and orbitals. Therefore, all  $D_{\vn,j}$ are actually constant $c$-numbers. Due to the anti-commutation relations of Majorana fermions, one can see that $D_{\vn,j}^2=1$ and thus $D_{\vn,j}=\pm1$. When the distribution of $D_{\vn,j}$ is given, the Hubbard interaction term becomes quadratic in terms of fermion operators, which makes the interacting model solvable.

Among all possible the $D_{\vn,j}$ configurations, there are two special ones with the potential to give rise the ground state of the interacting topological superconductor. The first one is the uniform distribution with $D_{\vn,j}=1$ for all sites and orbitals, which is called the ferromagnetic (FM) configuration. The other one is the staggered distribution with $D_{\vn,1}=1$ and $D_{\vn,2}=-1$ for all sites, which is called anti-ferromagnetic (AFM) configuration.

\begin{figure}
\centering
\includegraphics[width=0.45\textwidth]{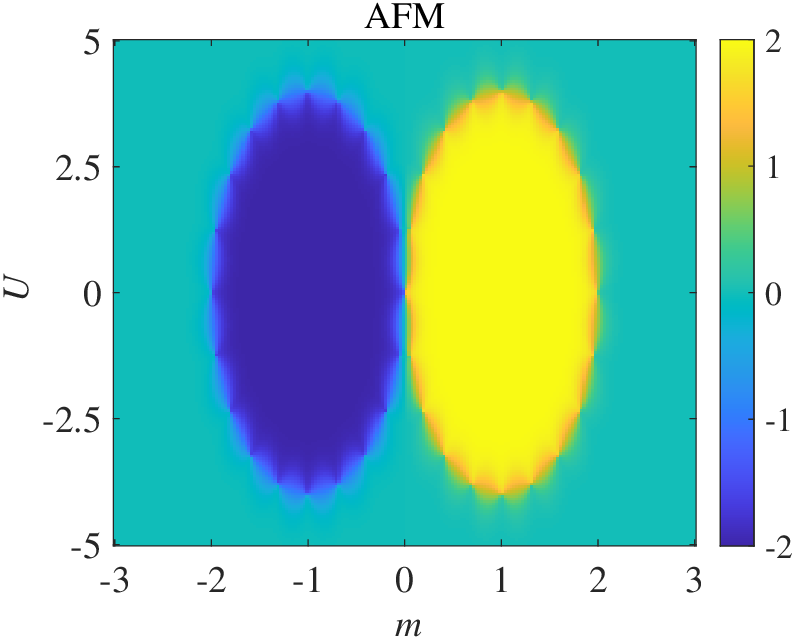}
\includegraphics[width=0.45\textwidth]{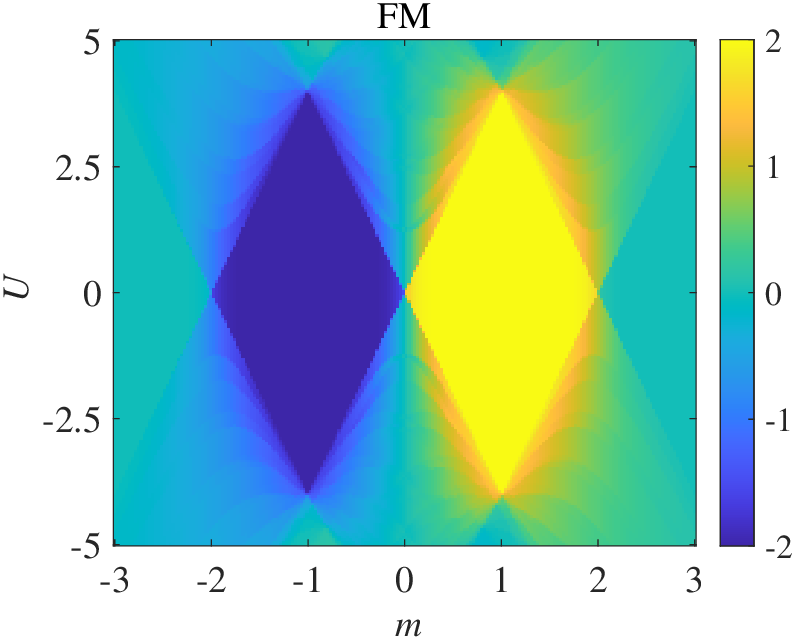}
\caption{The phase diagram of the interacting topological superconductor of Eq.(\ref{H-full}) with AFM configurations (left panel) and  FM configuration (right panel). The yellow and blue regions correspond the Chern number $C=\pm2$. The green region corresponds to $C=0$. The system size is $L_{x}=L_{y}=20$.}
\label{fig:phase}
\end{figure}

If we transfer the Majorana fermions back to ordinary fermions, the Hubbard interaction term in the first quantized form for the FM and AFM configurations can be written as
\be
&&H_{\textrm{int}}^{\textrm{FM}}=-\frac{U}{4}(\Gamma_{00}-\Gamma_{13})\otimes s_{2}\label{h:fm}\\
&&H_{\textrm{int}}^{\textrm{AFM}}=-\frac{U}{4}(\Gamma_{03}-\Gamma_{10})\otimes s_{2}\label{h:afm}
\ee
With this result, the interacting topological superconductor of Eq.(\ref{H-full}) can be converted into a simple free fermion model, for which the Chern number can be easily computed. Accordingly, we map out the phase diagram of the interacting topological superconductor in Fig.\ref{fig:phase}. The topological non-trivial phase is roughly located inside the area of $m\in(-2,2)$ and $U\in(-4,4)$. The yellow and blue regions indicate the phase with Chern number $C=\pm2$, while the green region indicate the topological trivial phase with $C=0$. We would like to mention that the true ground state of the model is the AFM configuration.

In this paper, we will focus on the disorder effects of the interacting topological superconductor of Eq.(\ref{H-full}). We always assume that the flat band condition $\Delta=1$ is satisfied, such that the interacting model can be exactly solved. In order to keep the solvability of this interacting model, we impose the disorders either on the coefficient $m$ or the coupling constant $U$. More explicitly, we will make the following replacement $m\to m_{\vn}=m+\omega_\vn$ or $U\to U_{\vn}=U+\omega_\vn$. In other words, we consider two types of disorder models as follows
\be
&&H_{\textrm{dis1}}=H_{\textrm{full}}+\omega_\vn(\Gamma_{31}-\Gamma_{22})\otimes s_{0},\label{dis-m}\\
&&H_{\textrm{dis2}}=H_{\textrm{full}}-\frac{\omega_\vn}{4}(\Gamma_{03}-\Gamma_{10})\otimes s_{2}
\label{dis-U}
\ee
Here $\omega_{\vn}$ located at each lattice site are random variables with uniform distribution inside the interval $[-W/2,W/2]$. $W$ is the disorder strength. We have numerically checked that the AFM configuration is also the true ground state even when the disorders are introduced. Therefore, the disorder potential term in Eq.(\ref{dis-U}) takes the form of Eq.(\ref{h:afm}).

\section{Numerical methods for disordered systems}
\label{method}
In this section, we discuss two different methods to investigate the topology of the interacting TS model with disorders. Usually, the Chern number is defined as an integral in the momentum space, but this form is not suitable for disordered system. Here we introduce a Chern number formula in real space, which can compute this topological index to high precision.
The other method is the so-called self-consistent Born approximation, which can convert the disorder system into a uniformed system with renormalized parameters. Then, it is convenient to understand how the disorder affects topological phase transitions.

The Chern number formula defined in $\vk$-space can be written as
\be
C=\frac{1}{2\pi i}\int d^{2}k\,\textrm{Tr}\Big(P_{\vk}[\partial_{x}P_{\vk},\partial_{y}P_{\vk}]\Big)
\ee
Here $P_{\vk}$ is the projection operator in the $\vk$-space, which is define as
\be
P_{\vk}=\sum_{E_{n}<E_{f}}\ket{\psi_{n}(\vk)}\bra{\psi_{n}(\vk)}
\label{pro:p}
\ee
where $\ket{\psi_{n}}$ represents the $n$-th eigenvector with the eigen-energy $E_n$, and $E_f$ represents the Fermi energy.
This operator $P_{\vk}$ projects the wave-function into the subspace spanned by the filled energy bands.

For the disordered systems, it is not possible to directly make $\vk$-space calculations. Therefore, one must seek a method to compute the Chern number in real space. If we diagonalize the Hamiltonian in real space, we can also construct a projection operator $P$ in real space. Next, we can approximate the derivative by finite differences. Then we will arrived at a Chern number formula in real space. This approach was suggested a few years ago by Prodan \cite{prodanPRL2010,Prodan_2011}. We will provide a detailed derivation of this formula in the appendix \ref{sec:chern}. In this section, we simply give out the real space Chern number formula as follows
\be
C=-\frac{1}{\pi}\textrm{Im Tr}\Big[P(K_{x}\cdot P)(K_{y}\cdot P)\Big]
\label{che:C}
\ee
Here the dot means element-wise product of the two matrices. ``Im'' means taking the imaginary part of the following quantities. The matrices $K_x$ and $K_y$ are given in the appendix \ref{sec:chern}.

Now we turn to the self-consistent Born approximation which is widely used in the many works such as \cite{chenrui2017PRB,chen2017disorderPRB,klier2019weak}. This method usually provide a useful way to understand topological  phase transitions driven by disorders. The essence of this method is to incorporate the effects of the disorder into a self-energy. Then the self-energy will in turn modify the parameters of the original model. Now the disordered model is actually converted into a uniform model with renormalized parameters. Then we can determine the boundary of topological phase transition by the phase diagram of Fig. \ref{fig:phase} with renormalized $m$ or $U$. In other words, the way that renormalized $m$ or $U$ depends on the disorder strength $W$ directly reflect the influence of disorders on topological phase transitions.

In the self-consistent Born approximation, the self-energy $\Sigma$ is defined as \cite{shen-book}
\be
(E_{f}-H_{\textrm{full}}(\vk)-\Sigma)^{-1}=\langle(E_{f}-H_{\textrm{full}}(\vk)-V_{\vn})^{-1}\rangle_{\textrm{dis}}
\label{func:green}
\ee
Here $H_{\textrm{full}}(\vk)$ is the interacting TS model of Eq.(\ref{H-full}) in the momentum space. $V_\vn$ is the disorder potential in real space. The $\langle\cdots\rangle_{\textrm{dis}}$ denotes the disorder average. For the two disorder models of Eq.(\ref{dis-m}) and (\ref{dis-U}), the disorder potential can be written as
\be
&&V_\vn=\omega_\vn \Lambda_s,\quad (s=1,2)\nonumber\\
&&\Lambda_1=(\Gamma_{31}-\Gamma_{22})\otimes s_0,\quad \Lambda_2=-\frac14(\Gamma_{03}-\Gamma_{10})\otimes s_2
\ee
where $\omega_\vn$ is the random number we defined before.

If we expand Eq.(\ref{func:green}) to the first order, we find that the disorder-induced self-energy is determined by the equation  \cite{GrothPRL2009,zhoubinPhysRevB2019} as follows
\be
\Sigma=\frac{W^{2}}{12}\frac{1}{(2\pi)^2}\int_{\textrm{BZ}}d^2k\,\Lambda_s[E_F-H_{\textrm{full}}(\vk)-\Sigma]^{-1}\Lambda_s
\label{eq:Born}
\ee
Here $E_F$ is the Fermi energy which is close to zero in our model. The integration is over the whole Brillouin zone.

For the disorder model of Eq.(\ref{dis-m}), by iterating Eq.(\ref{eq:Born}), we can solve the self-energy and find that $\Sigma=\delta m\,\Lambda_1$ plus some constant matrices which can be ignored. Now the effective Hamiltonian is $H=H_{\textrm{full}}+\Sigma$. Therefore, one can see that the self-energy only renormalize the parameter $m$  to $\widetilde{m}$, which is given by
\be
\widetilde{m}=m+\delta m
\ee
Similarly, if we consider the disorder model of Eq.(\ref{dis-U}), we find that the self-energy $\Sigma=\delta U\,\Lambda_2$ and the coupling $U$ is renormalized to $\widetilde{U}=U+\delta U$.

\section{Numerical results}
\label{num}

\begin{figure}
\centering
\includegraphics[width=0.45\textwidth]{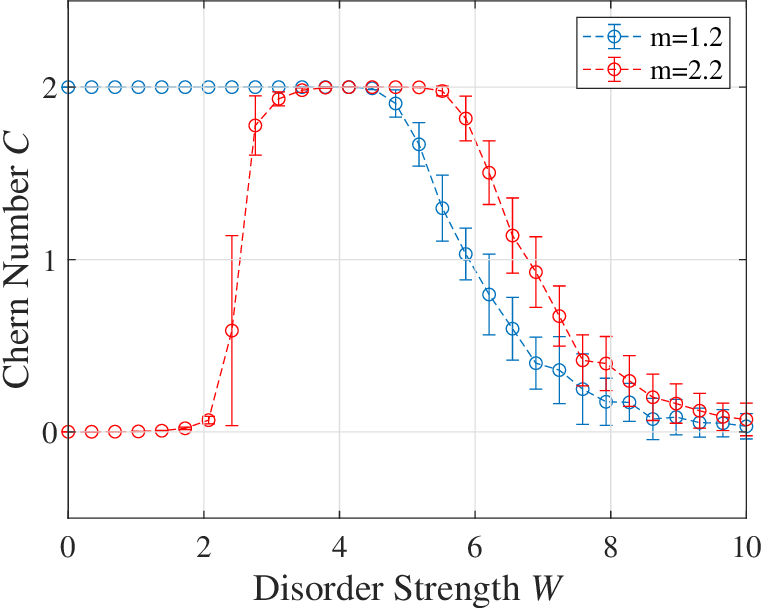}
\caption{The Chern number as a function of the disorder strength $W$ for $m=1.2$ (blue curve) and $m=2.2$ (red curve). The flat band condition $\Delta=1$ is assumed. The system size is $L_{x}=L_{y}=40$. The error bar indicates standard deviation of 30 random samples. The blue and red circles represent the values of Chern number.}
\label{fig:U0}
\end{figure}

In this section, we present the numerical results generated by the two methods that we discussed in the previous section. We start from with the real space Chern number for the disordered interacting topological superconductor. First, as a simple example, we consider the disorder model of Eq.(\ref{dis-m}) with $U=0$. In other words, we are working with Eq.(\ref{ham:eq4}) without the Hubbard interaction and the disorders are imposed on the potential term as in Eq.(\ref{dis-m}). Here we study the disorder effects on two different cases with $m=1.2$ or $m=2.2$ corresponding to topological or trivial phases of the clean TS model respectively. The Chern number as a function of disorder strength $W$ is shown in Fig. \ref{fig:U0}.

For the case of $m=1.2$ (blue curve), we have the Chern number $C=2$ in the clean limit. As the disorder strength $W$ increases, one can see that $C$ stays at value $2$ for a while and then start to drop at about $W=4.8$. It gradually decreases to zero at about $W=9$ and after that $C$ stays at zero. Therefore, there exists a critical disorder region where the topological phase transition takes place. This indicates that when the disorders are strong enough, they can kill the topological nontrivial phases.

For the other case of $m=2.2$ (red curve), the model is topologically trivial when $W=0$. However, when the disorder strength increased, one can see that the Chern number rises from $0$ to $2$ at around $W=1.7$ and stays at the value 2 for a while. It then starts to drop at $W=5.8$ and back to zero at around $W=8$. Within a certain range of disorder strength, a plateau of non-zero Chern number is maintained, which indicates that disorder can drive the system into a topologically nontrivial phase. This is very similar to the so-called topological Anderson insulator. But here it is realized in a topological superconductor instead of insulator.

\begin{figure}
\centering
\includegraphics[width=0.45\textwidth]{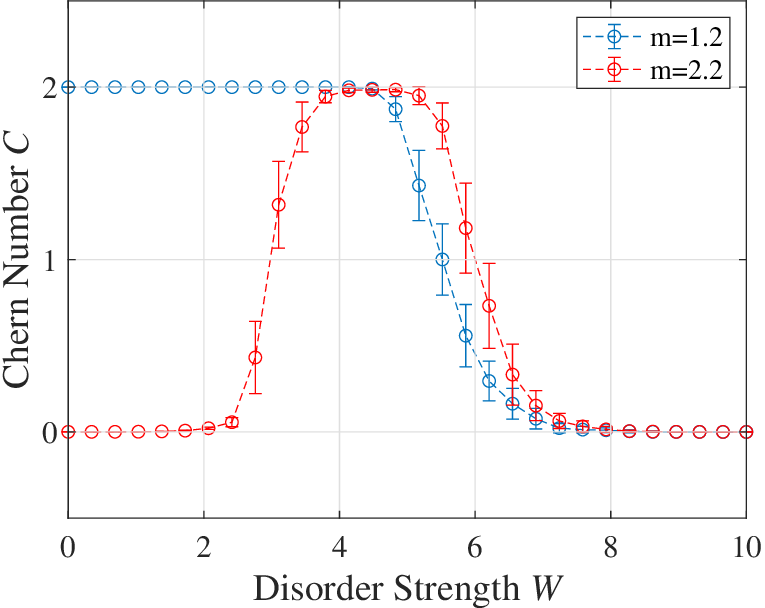}
\includegraphics[width=0.45\textwidth]{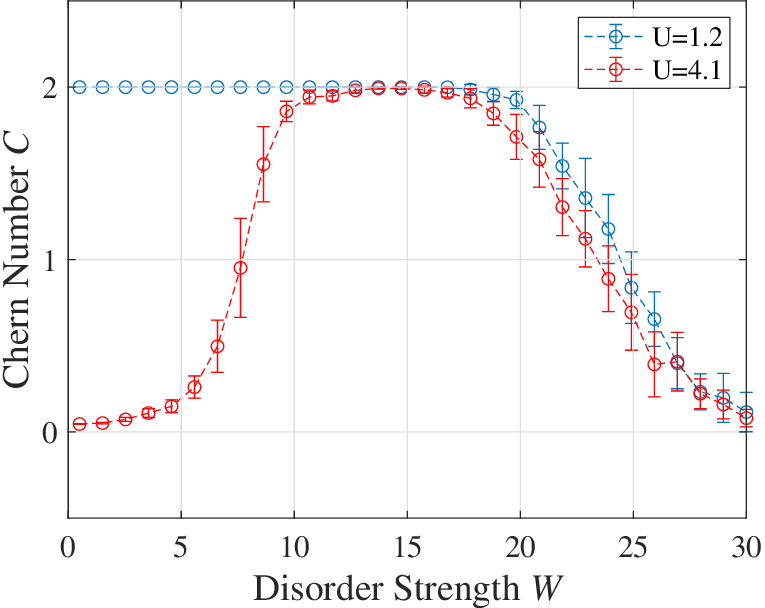}
\caption{The Chern number as a function of the disorder strength $W$. Left panel: The disorder model is Eq.(\ref{dis-m}) with $m=1.2,\,U=1.2$ (blue curve) and $m=2.2,\,U=1.2$ (red curve). Right panel: The disorder model is Eq.(\ref{dis-U}) with $m=1.2,\,U=1.2$ (blue curve) and $m=1.2,\,U=4.1$ (red curve). The system size is $L_{x}=L_{y}=40$. The error bars indicate standard deviation of 30 random samples.}
\label{fig:m-dis}
\end{figure}

Next we turn on the Hubbard interaction in the disorder model of Eq.(\ref{dis-m}). Since we assume $\Delta=1$, this mode is still exactly solvable. The result of its real space Chern number is presented in the left panel of Fig.\ref{fig:m-dis}.
Here we still consider two different cases with $m=1.2$ (blue curve) and $m=2.2$ (red curve). For both cases, we assume $U=1.2$. The behavior is very similar to the non-interacting case of Fig.\ref{fig:U0}. For $m=1.2$, the model is already topological for $W=0$. When $W$ increase to around $W=5$, the Chern number $C$ gradually drop to zero.  One the other hand, the model is trivial for $m=2.2$ in the clean limit. But the disorders will drive $C$ to increase from zero to $2$ and then return to zero again. The plateau of $C=2$ within a certain range of $W$ suggests that the disorder can induce a topologically nontrivial phase even when the Hubbard interaction is presented. This result generalizes the phenomena of topological Anderson transition to interacting models.

Since the interacting TS model is exact solvable, we can also impose the disorders on the coupling constant, as the disorder model of Eq.(\ref{dis-U}). The Chern number of this model is shown in the right panel of Fig. \ref{fig:m-dis}. The blue and red curve corresponds to the $U=1.2$ and $U=4.1$ respectively. In both cases, we assume $m=1.2$. One can see that the $U=1.2$ system has $C=2$ in the clean limit and $C$ gradually decreasing to zero inside the region of $20<W<30$.
According to the phase diagram of Fig.\ref{fig:phase}, the system with $U=4.1$ is topologically trivial in the clean limit. Again, one can see that the disorders can generate a stable plateau of $C=2$ inside the region of $10<W<20$. This is again the signature of topology driven by disorders. In the model of Eq.(\ref{dis-U}), the disorders come from the random coupling constant, which is usually difficult to analyze in other types of models.

We also would like to mention that the error bars in both Fig. \ref{fig:U0} and \ref{fig:m-dis} represent the standard deviations of 30 random samples. The error bars in the transition regions are larger than the ones in the stable regions.

\begin{figure}
\centering
\includegraphics[width=0.4\textwidth]{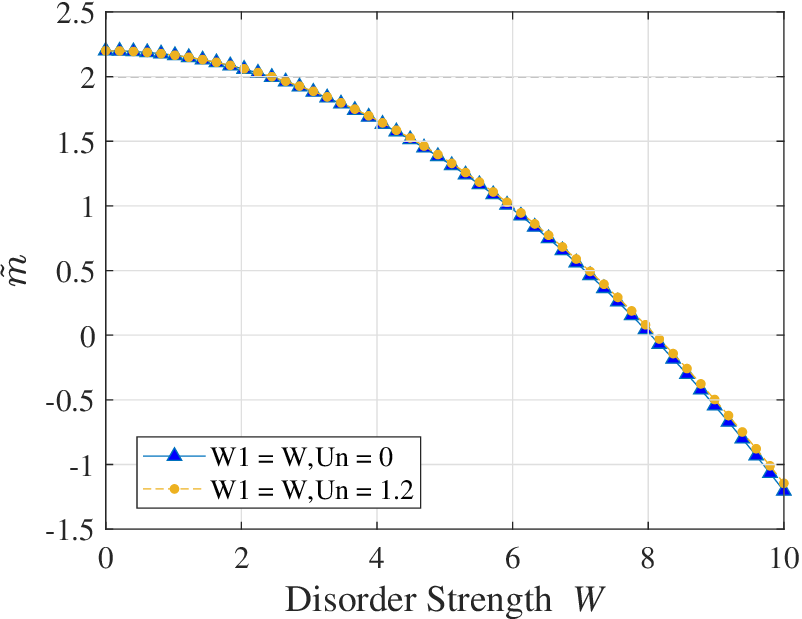}
\includegraphics[width=0.4\textwidth]{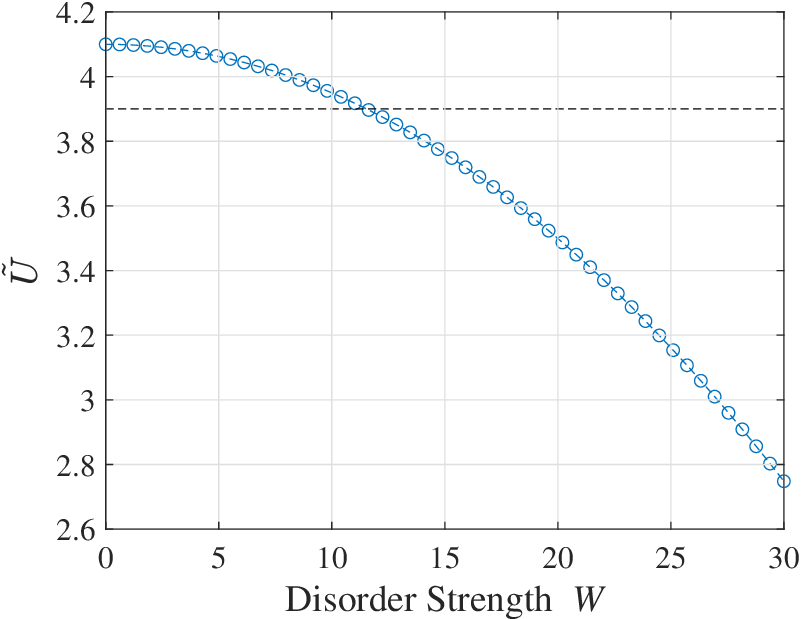}
\includegraphics[width=0.4\textwidth]{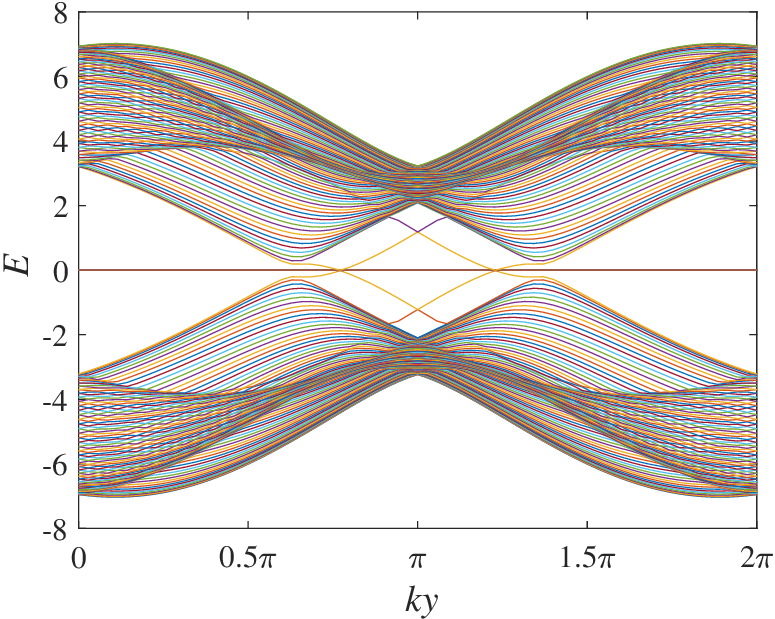}
\caption{Top left: The renormalized $\widetilde{m}$ as a function of the disorder strength $W$ with disorder of Eq.(\ref{dis-m}). We assume $m=2.2,\,U=0$ (blue triangle) and $m=2.2,\,U=1.2$ (yellow dot). Top right: The renormalized $\widetilde{U}$ as a function of $W$ with disorder of Eq.(\ref{dis-U}). Here we assume $m=1.2$. Bottom: The energy bands of Eq.(\ref{H-full}) with renormalized  $\widetilde{U}$ as a function of $k_{y}$. The parameters are $m=1.2, U=4.1, W=11$.}
\label{fig:Born}
\end{figure}

At last, we present the numerical results of effective medium theory based on the self-consistent Born approximation.
We first consider the disorder model of Eq.({\ref{dis-m}}) which renormalizes the parameter $m$. The renormalized $\widetilde{m}$ as a function of $W$ is displayed in the top left panel of Fig.\ref{fig:Born}. One can see that for both $m=2.2,\,U=0$ (blue triangle) and $m=2.2,\,U=1.2$ (yellow dot) cases, the two renormalized $\widetilde{m}$ are almost identical. The dashed line represents the topological phase boundary of $\widetilde{m}=2$. The corresponding critical disorder strength is $W\approx2$, which is close the region where $C=0$ rises to $C=2$ in the left panel of Fig.\ref{fig:m-dis}. For larger $W$, we find that $-2<\widetilde{m}<2$, which suggest the system stays in the topological phase even when $W$ increases up to 10. On the other hand the Chern number curve in the left panel of Fig.\ref{fig:m-dis} shows that $C$ drop to zero at $W\approx6$. This contradiction shows that the self-consistent Born approximation only works for the relatively weak disorders. It cannot capture the correct topological features when $W$ is too strong.

Similarly, we can also study the disorder model of Eq.({\ref{dis-U}}) which renormalizes the coupling $U$. The resulting $\widetilde{U}$ is shown in the right panel of Fig.\ref{fig:Born}. The dashed line represents the phase boundary of $\widetilde{U}=3.9$. One can see the corresponding critical $W\approx10$ which is close to the first transition in the right panel of Fig.\ref{fig:m-dis}. Again, the renormalized $\widetilde{U}$ fail to capture the second transition at strong $W$.

Another advantage of the self-consistent Born approximation is that it can eliminate the disorders such that the model becomes a uniform one again. Because of this, we can put the model on a cylinder geometry with open boundary along $x$-axes and closed boundary along $y$-axes. For example, the disorder model of Eq.(\ref{dis-U}) is converted to Eq.(\ref{H-full}) with renormalized  $\widetilde{U}$. Now the Hamiltonian as a function of $k_y$ is ready to be diagonalized. The resulting band structure is shown in the bottom panel of Fig.\ref{fig:Born}. One can see the appears of the edge modes due to the non-trivial topology. It is also evident that the system preserves its flat bands even with the introduction of disorders, thereby maintaining its exact solvability. Similar conclusions can be drawn for the disorder model of Eq.(\ref{dis-m}).

\section{conclusion}
\label{conclusion}

We have studied the topological phase transitions of the disordered topological superconductor with Hubbard interaction where the disorders are either imposed on the potential term or the coupling constants. Throughout the whole paper, we stick to the flat-band condition which guarantees the exact solvability of this model even when disorders are included.  It is found that the strong enough disorders are able to suppress the topological phases. For topological trivial system in the clean limit, the disorders also induce a phase with non-zero Chern number. This can be thought as a generalization of the topological Anderson transitions to interacting models. The results of real space Chern number is also confirmed by the self-consistent Born approximation. In the later method, we also demonstrated the influence of disorders on the system parameters and band structures.

\begin{acknowledgments}
This work is supported by the National Natural Science Foundation of China under Grant No. 11874272.
\end{acknowledgments}

\appendix
\section{Derivation of Real space Chern Number}
\label{sec:chern}

We start from the Chern number formula in momentum space, which we repeated here as follows
\be
C=\frac{1}{2\pi i}\int d^{2}k\,\textrm{Tr}\Big(P_{\vk}[\p_{x}P_{\vk},\p_{y}P_{\vk}]\Big)
\label{Chern}
\ee
here $\p_\mu=\frac{\p}{\p k_{\mu}}$ with $\mu=x,y$. By discretizing the continuous model, the Brillouin zone of a 2D square lattice is divided into a $N\times N$ grid, with separation $h=2\pi/N$. The derivative of $k_x$ can be approximated as
\be
\p_{x}P_{\vk}\approx \delta P_{\vk}\equiv\sum_{m=1}^{Q}\frac{c_{m}}{2h}[P_{\vk+mh\hat{x}}-P_{\vk-mh\hat{x}}]
\label{eq:app}
\ee
Here $\hat{x}$ is a unit vector along $x$-axes. The coefficients $c_{m}$ are some weights to be determined later. If we take $Q=1$ and $c_1=1$, we get back to a simple symmetric difference. In order to improve the precision, we assume $Q=N/2$ where $N$ is the number of lattice sites along the $x$-axes.  We can fourier transform $P_{\vk}$ as
\be
P_{\vk}=\sum_{\vbr}b_{\vbr}e^{i \vk\cdot\vbr}
\label{eq:Pk}
\ee
Substituting Eq. (\ref{eq:Pk}) into Eq.(\ref{eq:app}) we find that
\be
&&\p_x P_{\vk}=i\sum_{\vbr}b_{\vbr}x\,e^{i \vk\cdot\vbr}\\
&&\delta P_{\vk}=\sum_{\vbr}b_{\vbr}e^{i \vk\cdot\vbr}\sum_{m=1}^{Q}\frac{c_{m}}{2h}(2i)\sin(m h x)
\ee
The approximation in Eq.(\ref{eq:app}) implies the following condition
\be
x\approx\sum_{m=1}^{Q}\frac{c_{m}}{2h}2\sin(m h x)
\label{x-app}
\ee
Making a Taylor expansion of the right hand side of the above equation, we find that the above condition becomes a set of linear equations
\be
\sum_{m}^{Q}c_{m}m=1,\quad\sum_{m}^{Q}c_{m}m^{2j-1}=0
\label{cm}
\ee
from which we can numerically solve the weights $c_m$ with $m=1,\cdots,Q$.

In the real space representation, the projector $P_{\vk}$ is replaced by its real space counterpart $P$. The momentum shift of $P_{\vk}$ can be achieved by the translational operator as
\be
P_{\vk+mh\hx}\to e^{-i mhX}P e^{i mhX}
\ee
Here $X=\textrm{diag}\{1,2,\,\cdots,\,N\}\otimes I_{N}$ with the identity matrix $I_{N}$ of dimension $N$.
Now the approximated derivative $\delta P_{\vk}$ in the real space representation can be written as
\be
&&\delta P_{\vk}\to \sum_{m=1}^Q\frac{c_m}{2h}\Big(e^{-i mhX}P e^{i mhX}-e^{i mhX}P e^{-i mhX}\Big)_{ab}\nonumber\\
&&\quad=2i\sum_{m=1}^{Q}\frac{c_{m}}{2h}\sin mh(x_a-x_b)P_{ab}
\ee
Here $x_a$ is the diagonal elements of matrix $X$ . In summary, we find that the
\be
\p_x P_{\vk}\to \frac{i}{h} K_x\cdot P,\quad
(K_x)_{ab}=\sum_{m=1}^{Q}c_{m}\sin mh(x_a-x_b)
\ee
where the dot means element-wise matrix product. Similarly, we also have
\be
\p_y P_{\vk}\to \frac{i}{h} K_y\cdot P,\quad
(K_y)_{ab}=\sum_{m=1}^{Q}c_{m}\sin mh(y_a-y_b)
\ee
Here $y_a$ is the diagonal element of matrix $Y=I_{N}\otimes \textrm{diag}\{1,2,\,\cdots,\,N\}$.

Plug the above two equations into Eq.(\ref{Chern}), and notice that
\be
\textrm{Tr}\Big[P(K_{x}\cdot P)(K_{y}\cdot P)\Big]
=\textrm{Tr}\Big[P(K_{y}\cdot P)(K_{x}\cdot P)\Big]^\dag
\ee
we finally arrived at the following Chern number formula
\be
C=-\frac{1}{\pi}\textrm{Im Tr}\Big[P(K_{x}\cdot P)(K_{y}\cdot P)\Big]
\ee
which is the claimed result of Eq.(\ref{che:C}).


\end{document}